\documentclass[aps,floatfix]{revtex4}

\usepackage{graphicx}
\usepackage{amsmath}
\usepackage{amssymb}

\begin{document}

\title{Quantum dice rolling}

\author{N. Aharon}

\author{J. Silman}
\affiliation{School of Physics and Astronomy, Tel-Aviv University, Tel-Aviv 69978,
Israel}
\begin{abstract}\textbf{
A coin is just a two sided dice. Recently, Mochon proved that quantum
weak coin flipping with an arbitrarily small bias is possible. However,
the use of quantum resources to allow $N$ remote distrustful parties
to roll an $N$-sided dice has yet to be addressed. In this paper
we show that contrary to the classical case, $N$-sided dice rolling
with arbitrarily small bias is possible for any $N$. In addition,
we present a six-round three-sided dice rolling protocol, achieving a bias of $0.181$, which
incorporates weak imbalanced coin flipping.}
\end{abstract}
\maketitle
\section{Introduction}

Coin flipping is a cryptographic problem in which a pair of remote
distrustful parties, usually referred to as Alice and Bob, must generate
a random bit that they agree on. There are two types of coin flipping
protocols. In weak coin flipping (WCF) one of the parties prefers
one of the outcomes and the other prefers the opposite, whereas in
strong coin flipping (SCF) each party does not know the other's preference.
The security of a WCF protocol is quantified by the biases $\epsilon_{A}$
and $\epsilon_{B}$; if $P_{A}^{*}$ and $P_{B}^{*}$ are the maximal
winning probabilities achievable by a dishonest Alice and Bob, then
\begin{equation}
\epsilon_{i}\hat{=}P_{i}^{*}-\frac{1}{2}\,,\qquad i=A,\, B\,.\end{equation}
The biases tell us to what extent each of the parties can increase
their chances of winning beyond one half. $\epsilon\hat{=}\max\left(\epsilon_{A},\,\epsilon_{B}\right)$
is often referred to as the bias of the protocol. Correspondingly,
in SCF there are four biases.\\

The problem of coin flipping was first introduced by Blum in 1981,
who analyzed it in classical settings \cite{Blum}. It was subsequently
shown that if there are no limitations on the parties' computational
power a dishonest party can always force any outcome they desire \cite{Kilian}.
With the publication of the quantum key distribution protocol of Bennett
and Brassard in 1984\cite{BB84}, it was realized that many communication
tasks that are impossible in a classical setting may be possible in
a quantum setting. In 1999 Goldenberg \emph{et al.} introduced a quantum
gambling protocol \cite{Gambling}, which is a problem closely related
to WCF (see \cite{footnote}). The first quantum (strong) coin flipping
protocol per se was presented by Aharonov \emph{et al.} in 2000 \cite{Aharonov}.
The protocol achieves a bias of $\sqrt{2}/4\simeq0.354$ \cite{Spekkens -2}.
Soon afterward Spekkens and Rudolph \cite{Spekkens -1}, and independently
Ambainis \cite{Ambainis}, devised a SCF protocol with a bias
of $1/4$. On the other hand, Kitaev subsequently proved that there
is a limit to the efficacy of SCF protocols \cite{Kitaev}: Any SCF
protocol must satisfy $\left.P_{A}^{\left(i\right)}\right.^{*}\cdot\left.P_{B}^{\left(i\right)}\right.^{*}\geq1/2$,
$i\in\left\{ 0,\,1\right\} $, where $\left.P_{A}^{\left(i\right)}\right.^{*}$
($\left.P_{B}^{\left(i\right)}\right.^{*}$) is dishonest Alice's
(Bob's) maximal probability of forcing the outcome $i$. As regards
WCF, in 2002 Spekkens and Rudolph introduced a family of three rounds
of communication protocols in which both dishonest parties have a
bias of $(\sqrt{2}-1)/2\simeq0.207$ \cite{Spekkens}. Mochon then
improved upon Spekkens and Rudolph's result by constructing WCF protocols
with an infinite number of rounds \cite{Mochon_-2,Mochon_-1}. These
efforts culminated in a proof that WCF with an arbitrarily small bias
is possible \cite{Mochon}. Most recently, building upon Mochon's
latest result, Chailloux and Kerenidis devised a SCF protocol, which
saturates Kitaev's bound in the limit of an infinite number of rounds
\cite{Kerenidis}.\\

The paper is constructed as follows. In section II we define the problem
of dice rolling and prove that, using quantum resources, dice rolling
with an arbitrarily small bias is possible. This result stands in
marked contrast to the classical case, where, under certain conditions,
an honest party always loses. In section III we analyze a three-round
weak imbalanced coin flipping protocol, which generalizes the results of Spekkens and Rudolph  
to the imbalanced case.
The protocol is then used in section IV to implement a six-round three-sided
dice rolling protocol; the motivation being to provide an indication
of the degree of security achievable by a protocol with a minimal
number of rounds of communication. We conclude with a short discussion.

\section{Quantum dice rolling with arbitrarily small bias}

It is straightforward to generalize the problem of coin flipping to
multi-party settings. In multi-party coin flipping $N$ remote distrustful
parties must decide on a bit. An analysis of multi-party SCF in a
quantum setting was carried out in \cite{Multi-party}, and similarly
to the two-party case, it was shown that the use of quantum resources
is advantageous. Another possible generalization is the problem of
$N$ remote distrustful parties having to agree on a number from $1$
to $N$, with party $i$ preferring the $i$-th outcome. To the best
of our knowledge this problem has never been considered in classical
settings. We shall term this problem {}``dice rolling''. For $N=2$
the problem reduces to WCF, while for $N>2$, there are now many different
cheating scenarios, as any number of parties $n<N$ may be dishonest.
We will be interested in the $N$ {}``worst case'' scenarios where
all but one of the parties are dishonest and, moreover, are cooperating
with one another. That is, the dishonest parties share classical and
quantum communication channels. In addition, we will require that
the protocol be {}``fair'' in the sense that the honest party's
maximum losing probability be the same in each of these $N$ scenarios.
Of course, the security of the protocol can be evaluated with respect
to any other cheating scenario, but as we will consider only fair
protocols, the security of any cheating scenarios is never poorer
than that provided by the aforementioned $N$ scenarios.\\

We begin by observing that in coin flipping protocols the bias has
a complementary definition. We could just as well define it as \begin{equation}
\bar{\epsilon}_{i}\hat{=}\bar{P}_{i}^{*}-1/2\,,\qquad i=A,\, B\end{equation}
where $\bar{P}_{i}^{*}=P_{j\neq i}^{*}$ is the maximum probability
that party $i$ loses. According to this definition the bias tells
us to what extent party $j\neq i$ can increase the other party $i$'s
chances of losing beyond one half. In the case of $N$ parties, the
bias $\bar{\epsilon}_{i}$ then tells us to what extent the $N-1$
dishonest parties can increase party $i$'s chances of losing beyond
$1-1/N$, rather than to what extent a sole dishonest party can increase
its chances of winning beyond $1/N$. We shall always use this redefinition
of the bias when considering dice rolling. The computation of biases
in dice rolling is therefore equivalent to the computation of biases
in a weak imbalanced coin flipping protocol.

We shall now prove that that quantum weak $N$-sided dice rolling
with arbitrarily small bias is possible for any $N$. The proof is
by construction. Consider the following $N$-party protocol. Each
party is uniquely identified according to a number from $1$ to $N$.
The protocol consists of $N-1$ stages. In stage one parties $1$
and 2 {}``weakly flip'' a balanced quantum coin. The winner and
party $3$ then weakly flip an imbalanced quantum coin in stage two,
where if both parties are honest $3$'s winning probability equals
$1/3$. And so on, the rule being that in stage $n\geq2$ the winner
of stage $n-1$ and party $n$ {}``weakly flip'' an imbalanced quantum
coin, where if both parties are honest, $n$'s winning probability
equals $1/\left(n+1\right)$. Thus, if all parties are honest each
has the same overall winning probability of $1/N$. Using Mochon's
formalism \cite{Mochon}, Chailloux and Kerenidis have recently proved
that weak imbalanced coin flipping with arbitrarily small bias is
possible \cite{Kerenidis}. It follows that in the limit where each
of the weak imbalanced coin flipping protocols, used to implement
our dice rolling protocol, admits a vanishing bias (and $N$ is finite),
any honest party's winning probability tends to $1/N$; for a formal
proof see the appendix. Moreover, since we have considered the worst
case cheating scenario, this result holds for any other cheating scenario.

The above result is in stark contrast to the classical case where
if the number of honest parties is not strictly greater than $N/2$,
then the dishonest parties can force any outcome they desire. To see
why this is so, let us consider a classical $N$-sided dice rolling
protocol and partition the parties into two groups of $m\le\left\lceil N/2\right\rceil $
and $n=N-m$ parties. If both groups are honest, the probability that
a party in the first (second) group wins is $m/N$ ($1-m/N$). Therefore,
any dice rolling protocol can serve as a weak imbalanced coin flipping
protocol. Suppose now that all of the parties in the second group
are dishonest, and are nevertheless unable to force with certainty
the outcome they choose. Clearly, this would still be the case even
if they were the smaller group, i.e. $n<m$ ($m>\left\lceil N/2\right\rceil $),
and we get a contradiction, since in classical weak imbalanced coin
flipping (as in weak balanced coin flipping) at least one of the parties
should always be able to force whichever outcome they desire \cite{Keyl}.

\section{A three-round weak imbalanced coin flipping protocol}

In this section we analyze a three-round weak imbalanced coin
flipping protocol based on quantum gambling. It is constructed such
that if both parties are honest Alice's winning probability equals
$1-p$. Interestingly, it turns out that this protocol coincides with the generalization of Spekkens and Rudolph's work to
the imbalanced case.
The protocol will be used in the subsequent section to implement
a six-round three-sided dice rolling protocol.

\subsection*{The protocol}
\begin{enumerate}
\item Alice prepares a superposition of two qubits \begin{equation}
\left|\psi_{0}\right\rangle =\sqrt{1-p-\eta}\left|\uparrow_{1}\downarrow_{2}\right\rangle +\sqrt{p+\eta}\left|\downarrow_{1}\uparrow_{2}\right\rangle ,\qquad0\leq\eta\leq1-p\,,\end{equation}
where the subscripts serve to distinguish between the first and second
qubit and will be omitted when the distinction is clear. She then
sends the second qubit to Bob.
\item Bob carries out a unitary transformation $U_{\eta}$ on the qubit
he received and another qubit (labelled by the subscript $3$) prepared
in the state $\left|\downarrow\right\rangle $ such that \begin{equation}
\left|\uparrow_{2}\downarrow_{3}\right\rangle \rightarrow U_{\eta}\left|\uparrow_{2}\downarrow_{3}\right\rangle =\sqrt{\frac{p}{p+\eta}}\left|\uparrow_{2}\downarrow_{3}\right\rangle +\sqrt{\frac{\eta}{p+\eta}}\left|\downarrow_{2}\uparrow_{3}\right\rangle \,,\end{equation}
and \begin{equation}
\left|\downarrow_{2}\uparrow_{3}\right\rangle \rightarrow U_{\eta}\left|\downarrow_{2}\uparrow_{3}\right\rangle =\sqrt{\frac{\eta}{p+\eta}}\left|\uparrow_{2}\downarrow_{3}\right\rangle -\sqrt{\frac{p}{p+\eta}}\left|\downarrow_{2}\uparrow_{3}\right\rangle \,,\end{equation}
with $U_{\eta}$ acting trivially on all other states. The resulting
state is then \begin{equation}
\left|\psi_{1}\right\rangle =U_{\eta}\left|\psi_{0}\right\rangle =\sqrt{1-p-\eta}\left|\uparrow_{1}\downarrow_{2}\downarrow_{3}\right\rangle +\sqrt{p}\left|\downarrow_{1}\uparrow_{2}\downarrow_{3}\right\rangle +\sqrt{\eta}\left|\downarrow_{1}\downarrow_{2}\uparrow_{3}\right\rangle \,.\end{equation}
Following this, he checks whether the second and third qubits are
in the state $\left|\uparrow_{2}\downarrow_{3}\right\rangle $.
\item Bob wins if he finds the qubits in the state $\left|\uparrow_{2}\downarrow_{3}\right\rangle $.
Alice then checks whether the first qubit is in the state $\left|\downarrow\right\rangle $,
in which case Bob passes the test. If Bob does not find the qubits
in the state $\left|\uparrow_{2}\downarrow_{3}\right\rangle $, he
asks Alice for the first qubit and checks whether all three qubits
are in the state \begin{equation}
\left|\xi\right\rangle \hat{=}\sqrt{\frac{1-p-\eta}{1-p}}\left|\uparrow_{1}\downarrow_{2}\downarrow_{3}\right\rangle +\sqrt{\frac{\eta}{1-p}}\left|\downarrow_{1}\downarrow_{2}\uparrow_{3}\right\rangle \,,\end{equation}
in which case she passes the test.
\end{enumerate}

\subsection*{Alice's maximal bias}

Most generally Alice can prepare any state of the form \begin{equation}
\left|\psi_{0}'\right\rangle =\sum_{i,\, j=\uparrow,\,\downarrow}\alpha_{ij}\left|ij\right\rangle \otimes\left|\Phi_{ij}\right\rangle \,,\end{equation}
where the $\left|\Phi_{ij}\right\rangle $ are states of some ancillary
system at her possession. After Bob applies $U_{\eta}$ the resulting
composite state is given by \begin{eqnarray}
 & \left|\psi_{1}'\right\rangle =U_{\eta}\left|\psi_{0}'\right\rangle \otimes\left|\downarrow\right\rangle =\alpha_{\uparrow\uparrow}\left(\sqrt{\frac{p}{p+\eta}}\left|\uparrow\uparrow\downarrow\right\rangle +\sqrt{\frac{\eta}{p+\eta}}\left|\uparrow\downarrow\uparrow\right\rangle \right)\otimes\left|\Phi_{\uparrow\uparrow}\right\rangle \\
 & +\alpha_{\uparrow\downarrow}\left|\uparrow\downarrow\downarrow\right\rangle \otimes\left|\Phi_{\uparrow\downarrow}\right\rangle +\alpha_{\downarrow\uparrow}\left(\sqrt{\frac{p}{p+\eta}}\left|\downarrow\uparrow\downarrow\right\rangle +\sqrt{\frac{\eta}{p+\eta}}\left|\downarrow\downarrow\uparrow\right\rangle \right)\otimes\left|\Phi_{\downarrow\uparrow}\right\rangle +\alpha_{\downarrow\downarrow}\left|\downarrow\downarrow\downarrow\right\rangle \otimes\left|\Phi_{\downarrow\downarrow}\right\rangle \,.\nonumber \end{eqnarray}
The probability that Bob does not find find the second and third qubits
in the state $\left|\uparrow_{2}\downarrow_{3}\right\rangle $ is
\begin{equation}
\bar{P}_{\uparrow\downarrow}=1-P_{\uparrow\downarrow}=1-\frac{\left|\alpha_{\uparrow\uparrow}\right|^{2}p+\left|\alpha_{\downarrow\uparrow}\right|^{2}p}{p+\eta}\,,\end{equation}
and the resulting composite state then is \begin{eqnarray}
 & \left|\psi_{2}'\right\rangle =\mathcal{N}\left(\alpha_{\uparrow\uparrow}\sqrt{\frac{\eta}{p+\eta}}\left|\uparrow\downarrow\uparrow\right\rangle \otimes\left|\Phi_{\uparrow\uparrow}\right\rangle +\alpha_{\uparrow\downarrow}\left|\uparrow\downarrow\downarrow\right\rangle \otimes\left|\Phi_{\uparrow\downarrow}\right\rangle \right.\nonumber \\
 & \left.+\alpha_{\downarrow\uparrow}\sqrt{\frac{\eta}{p+\eta}}\left|\downarrow\downarrow\uparrow\right\rangle \otimes\left|\Phi_{\downarrow\uparrow}\right\rangle +\alpha_{\downarrow\downarrow}\left|\downarrow\downarrow\downarrow\right\rangle \otimes\left|\Phi_{\downarrow\downarrow}\right\rangle \right)\,,\quad\end{eqnarray}
where $\mathcal{N}$, the normalization, is \begin{equation}
\frac{1}{\mathcal{N}^{2}}=1-\frac{p}{p+\eta}\left(\left|\alpha_{\uparrow\uparrow}\right|^{2}+\left|\alpha_{\downarrow\uparrow}\right|^{2}\right)\,.\end{equation}
The probability that Alice passes the test is therefore given by \begin{equation}
P_{\mathrm{test}}=\left\Vert \left\langle \xi\mid\psi_{2}'\right\rangle \right\Vert ^{2}=\mathcal{N}^{2}\left\Vert \alpha_{\uparrow\downarrow}\sqrt{\frac{1-p-\eta}{1-p}}\left|\Phi_{\uparrow\downarrow}\right\rangle +\alpha_{\downarrow\uparrow}\sqrt{\frac{\eta^{2}}{\left(1-p\right)\left(p+\eta\right)}}\left|\Phi_{\downarrow\uparrow}\right\rangle \right\Vert ^{2}\,.\end{equation}
The maximum obtains for $\left|\Phi_{\uparrow\downarrow}\right\rangle =\left|\Phi_{\downarrow\uparrow}\right\rangle $.
This choice of the ancillary states does not affect the maximum of
$\bar{P}_{\uparrow\downarrow}$. Hence, Alice obtains no advantage
by using ancillary systems and we can do away with them. Alice's maximum
cheating probability is then \begin{equation}
P_{A}^{*}=\max_{\alpha_{ij}}\bar{P}_{\uparrow\downarrow}\cdot P_{\mathrm{test}}\,,\end{equation}
where now \begin{equation}
\bar{P}_{\uparrow\downarrow}\cdot P_{\mathrm{test}}=\left|\alpha_{\uparrow\downarrow}\sqrt{\frac{1-p-\eta}{1-p}}+\alpha_{\downarrow\uparrow}\sqrt{\frac{\eta^{2}}{\left(1-p\right)\left(p+\eta\right)}}\right|^{2}\end{equation}
($\bar{P}_{\uparrow\downarrow}=1/\mathcal{N}^{2}$). Clearly, this
expression is maximum when $\alpha_{\uparrow\uparrow}=\alpha_{\downarrow\downarrow}=0$.
Therefore, to maximize her chance of successfully cheating Alice will
prepare a state of the form \begin{equation}
\left|\psi_{0}'\right\rangle =\sqrt{1-\delta}\left|\uparrow_{1}\downarrow_{2}\right\rangle +\sqrt{\delta}\left|\downarrow_{1}\uparrow_{2}\right\rangle \,,\end{equation}
where with no loss of generality we have set $\alpha_{\uparrow\downarrow}=\sqrt{1-\delta}$
and $\alpha_{\downarrow\uparrow}=\sqrt{\delta}$. So that \begin{equation}
P_{A}^{*}=\max_{\delta}\left(\sqrt{\frac{\left(1-p-\eta\right)\left(1-\delta\right)}{1-p}}+\sqrt{\frac{\eta^{2}\delta}{\left(1-p\right)\left(p+\eta\right)}}\right)^{2}\,.\end{equation}

\subsection*{Bob's maximal bias}

Bob wins and passes the test whenever Alice does not find the first
qubit in the state $\left|\uparrow\right\rangle $. The probability
for this is just $p+\eta$. This gives an upper bound on Bob's maximal
cheating probability, which is reached if Bob always announces that
he has won. That is, \begin{equation}
P_{B}^{*}=p+\eta\,.\end{equation}

\subsection*{Biases in the balanced case}

In the balanced case a protocol is fair if $P_{A}^{*}=P_{B}^{*}$.
We can play with $\eta$ to make $P_{A}^{*}$ and $P_{B}^{*}$ minimal
under this constraint. It is easy to show that the minimum then obtains
for $\eta=\left(\sqrt{2}-1\right)/2$. It follows that $\epsilon_{A}=\epsilon_{B}=\left(\sqrt{2}-1\right)/2$
and $P_{A}^{\mathrm{*}}=P_{B}^{*}=1/\sqrt{2}$.

\section{A six-round three-sided dice rolling protocol with a bias of 0.181}

Apart from the inherent limitations on the security of a multi-party
quantum cryptographic protocol, it is most interesting, both from a theoretical
and a practical viewpoint, to determine what degree of security is
afforded using the least amount of communication. In this section we
introduce a six-round three-sided dice rolling protocol following
the general construction presented in the second section. The protocol
consists of two three-round stages. In the first stage, we have Alice
and Bob flip a balanced quantum coin. Following this, in the second
stage, the winner and Claire flip an imbalanced quantum coin, such
that if both parties are honest Claire's winning probability equals
$1/3$. The protocol is considered fair if $\bar{P}_{A}^{*}=\bar{P}_{B}^{*}=\bar{P}_{C}^{*}$.
Due to the protocols' symmetry with respect to the interchange of
Alice and Bob there are only two nonequivalent worst case scenarios,
i.e. either only Alice is honest or only Claire is honest.

Using the protocol in the previous section an honest Alice has a maximum chance of $1-1/\sqrt{2}$
of progressing to the second stage. Therefore, Alice's maximum losing
probability is given by\begin{equation}
\bar{P}_{A}^{\mathrm{*}}=\frac{1}{\sqrt{2}}+\left(1-\frac{1}{\sqrt{2}}\right)\bar{\Pi}_{2/3}^{*}\,,\end{equation}
while an honest Claire's maximum losing probability is given by \begin{equation}
\bar{P}_{C}^{\mathrm{*}}=\bar{\Pi}_{1/3}^{*}\,,\end{equation}
with $\bar{\Pi}_{1/3}^{*}$ ($\bar{\Pi}_{2/3}^{*}$) the maximum losing
probability of the party with a winning probability of $1/3$ ($2/3$)
when both parties are honest. Hence, we require that \begin{equation}
\bar{\Pi}_{1/3}^{*}=\frac{1}{\sqrt{2}}+\left(1-\frac{1}{\sqrt{2}}\right)\bar{\Pi}_{2/3}^{*}\,.\end{equation}
If we use the WCF protocol of the previous section to implement the
second stage, then $\bar{\Pi}_{1/3}^{*}$ and $\bar{\Pi}_{2/3}^{*}$,
and hence the $\bar{P}_{i}^{*}$, will depend on $\eta$. We then
have to minimize the $\bar{P}_{i}^{*}$ with respect to $\eta$ under
the constraint that they are all equal, or what is the same thing,
minimize $\bar{\Pi}_{1/3}^{*}$ under the constraint eq. (20). However,
there are two possible implementations. Either $1-p=2/3$ and the
second stage begins with Alice preparing the state $\sqrt{2/3-\eta}\left|\uparrow_{1}\downarrow_{2}\right\rangle +\sqrt{1/3+\eta}\left|\downarrow_{1}\uparrow_{2}\right\rangle $,
or else $1-p=1/3$ and the second stage begins with Claire preparing
the state $\sqrt{1/3-\eta}\left|\uparrow_{1}\downarrow_{2}\right\rangle +\sqrt{2/3+\eta}\left|\downarrow_{1}\uparrow_{2}\right\rangle $.
In the first case we have to compute \begin{equation}
\min_{\eta}\max_{\delta}\frac{1}{2}\left(\sqrt{\left(2-3\eta\right)\left(1-\delta\right)}+\sqrt{\frac{9\eta^{2}\delta}{\left(1+3\eta\right)}}\right)^{2}\end{equation}
under the constraint that \begin{equation}
\max_{\delta}\frac{1}{2}\left(\sqrt{\left(2-3\eta\right)\left(1-\delta\right)}+\sqrt{\frac{9\eta^{2}\delta}{\left(1+3\eta\right)}}\right)^{2}=\frac{1}{\sqrt{2}}+\left(1-\frac{1}{\sqrt{2}}\right)\left(\frac{1}{3}+\eta\right)\,,\end{equation}
while in the second case we have to compute \begin{equation}
\min_{\eta}\left(\frac{2}{3}+\eta\right)\end{equation}
under the constraint that \begin{equation}
\left(\frac{2}{3}+\eta\right)=\frac{1}{\sqrt{2}}+\left(1-\frac{1}{\sqrt{2}}\right)\max_{\delta}\left(\sqrt{\left(1-3\eta\right)\left(1-\delta\right)}+\sqrt{\frac{9\eta^{2}\delta}{\left(2+3\eta\right)}}\right)\,.\end{equation}
The first of these yields the lower bias $\bar{\epsilon}_{A}=\bar{\epsilon}_{B}=\bar{\epsilon}_{C}\simeq0.181$
corresponding to $\bar{P}_{A}^{*}=\bar{P}_{B}^{*}=\bar{P}_{C}^{*}\simeq0.848$.
The second yields a bias of $0.199$.

\section{Summary and discussion}

We have defined and studied a novel mutli-party generalization of
weak coin flipping, which we have termed dice rolling. In $N$-sided
dice rolling $N$ remote distrustful parties must decide on a number
between $1$ and $N$, with party $i$ having the preference for outcome
$i$. We have shown that for any bias $\bar{\epsilon}\ll1$ there
exists a quantum protocol such that an honest party is always guaranteed
a $1/N-\bar{\epsilon}$ chance of winning. This is in contrast with
classical protocols where for $n\leq\left\lceil N/2\right\rceil $
honest parties, the dishonest parties can force any outcome they desire.
We have also presented a six-round three-sided dice rolling protocol
in which the honest party's maximum losing probability equals $0.848$
as compared to the $2/3$ that obtains when all parties are honest.
Finally, this problem also admits a "strong" variant in which $M \geq 2$ remote distrustful parties must agree on 
a number between $1$ and $N>2$, without any party being aware of any other's preference. Kitaev's bound generalizes to these cases. In partciular, for $M=2$
it can be saturated, as will be shown in an upcoming publication  \cite{QDR}.\\
\\
We thank Lev Vaidman and Oded Regev for useful comments. N. Aharon acknowledges the support of the
Wolfson Foundation. J. Silman acknowledges the support of the the
Israeli Science Foundation.

\appendix
\section{}

For any dice rolling protocol, based on weak imbalanced coin flipping
according to the scheme presented in section II, party $n$'s maximum
chance of losing is given by \begin{eqnarray}
 & \bar{P}_{n}^{*}=\frac{N-1}{N}+\bar{\epsilon}_{n}=\mbox{\ensuremath{\bar{\Pi}}}_{n-1}^{*}+\sum_{k=n}^{N-1}\mbox{\ensuremath{\bar{\Pi}}}_{k}^{*}\prod_{j=0}^{k-n}\left(1-\bar{\Pi}_{n-1+j}^{*}\right)\\
 & =\frac{n-1}{n}+\bar{\delta}_{n-1}+\sum_{k=n}^{N-1}\left(\frac{1}{k}+\bar{\delta}_{k}\right)\left(\frac{1}{n}-\bar{\delta}_{n-1}\right)\prod_{j=1}^{k-n-1}\left(\frac{n+j}{n+j+1}-\bar{\delta}_{n-1+j}\right)\,,\nonumber \end{eqnarray}
where $\bar{\Pi}_{k}^{*}$ is party $n$'s maximum chance of losing
stage $k$ conditional on having made it to that round and $\bar{\delta}_{k}$
the corresponding bias. If we now let $\bar{\delta}_{\mathrm{max}}^{\left(n\right)}\hat{=}\max_{k}\bar{\delta}_{k}$
and $\bar{\delta}_{\mathrm{min}}^{\left(n\right)}\hat{=}\min_{k}\bar{\delta}_{k}$
($k=n-1,\,\dots,\, N-1$), then 
\begin{eqnarray}
 & \bar{\epsilon}_{n}\leq\bar{\delta}_{\mathrm{max}}^{\left(n\right)}+\bar{\delta}_{+}^{\left(n\right)}\sum_{k=n}^{N-1}\left(\frac{1}{n}-\bar{\delta}_{\mathrm{min}}^{\left(n\right)}\right)\prod_{j=1}^{k-n-1}\left(\frac{n+j}{n+j+1}-\bar{\delta}_{\mathrm{min}}^{\left(n\right)}\right)\nonumber \\
 & -\bar{\delta}_{\mathrm{min}}^{\left(n\right)}\sum_{k=n}^{N-1}\left(\frac{1}{k}+\bar{\delta}_{\mathrm{max}}^{\left(n\right)}\right)\prod_{j=1}^{k-n-1}\left(\frac{n+j}{n+j+1}-\bar{\delta}_{\mathrm{min}}^{\left(n\right)}\right)\nonumber \\
 & -\bar{\delta}_{\mathrm{min}}^{\left(n\right)}\sum_{k=n}^{N-1}\left(\frac{1}{k}+\bar{\delta}_{\mathrm{max}}^{\left(n\right)}\right)\left(\frac{1}{n}-\bar{\delta}_{\mathrm{min}}^{\left(n\right)}\right)\sum_{m=1}^{k-n-1}\prod_{j\neq m}\left(\frac{n+j}{n+j+1}-\bar{\delta}_{\mathrm{min}}^{\left(n\right)}\right)\nonumber \\
 & <\bar{\delta}_{\mathrm{max}}^{\left(n\right)}+\bar{\delta}_{\mathrm{max}}^{\left(n\right)}\sum_{k=n}^{N-1}\frac{1}{n}\prod_{j=1}^{k-n-1}\frac{n+j}{n+j+1}\nonumber \\
 & <N\bar{\delta}_{\mathrm{max}}^{\left(n\right)}\end{eqnarray}
Hence, if each of the weak imbalanced coin flipping protocols, used
to implement the dice rolling protocol, are such that $\bar{\delta}_{\mathrm{max}}^{\left(n\right)}\ll1/N$
for any $n$, an honest party's winning probability tends to $1/N$.


\begin{thebibliography}{17}
\bibitem{footnote}Prior to Mochon's work \cite{Mochon}, there were
serious doubts as to the possibility of WCF with arbitrarily small
bias. Quantum gambling was introduced as an example of
a quantum cryptographic task, impossible in classical settings, which
is neveretheless unconditionally secure in the sense that it does
away with the notion of cheating; both the casino and the player do
not have to follow any ``honest'' protocol, but rather are allowed
to do anything to maximize their gain.
\end{thebibliography}
\end{document}